\documentclass[twocolumn,prc,showpacs,preprintnumbers,
               unsortedaddress,amsmath,amssymb,floatfix]{revtex4}

\usepackage{graphicx,epsfig,latexsym,amssymb}
\usepackage{multirow,amsmath,array,booktabs}
\usepackage{dcolumn}
\usepackage[section]{placeins}
\usepackage{bm}

\begin{document}

\hspace{14.0cm}\hfill {\tt CAS-KITPC/ITP-084}\\

\title{P Wave Meson Spectrum in a Relativistic Model with Instanton
Induced Interaction}
\author{Bhavyashri, K. B. Vijaya Kumar$^{*}$}
 \affiliation{Department of Physics, Mangalore University,
              Mangalagangotri, Mangalore 574 199, India}

 \email{kbvijayakumar@yahoo.com}

\author{Yong-Liang Ma}
 \email{ylma@itp.ac.cn}
 \affiliation{Kavli Institute for Theoretical Physics China, Chinese Academy of Science, Beijing 100190, China}

\author{Antony Prakash}
 \affiliation{Department of Physics, Mangalore University,
              Mangalagangotri, Mangalore 574 199, India}

\date{\today}


\begin{abstract}
On the basis of the phenomenological relativistic harmonic models
for quarks we have obtained the masses of P wave mesons. The full
Hamiltonian used in the investigation has Lorentz scalar + vector
confinement potential, along with one gluon exchange potential
(OGEP) and the instanton-induced quark-antiquark interaction (III).
A good agreement is obtained with the experimental masses. The
respective role of III and OGEP for the determination of the meson
masses is discussed.
\end{abstract}

\maketitle

\section{Introduction}
\label{sec:intro}

There is a wealth of experimental data in hadron spectroscopy that
would constitute a good testing ground for nonperturbative quantum
chromodynamics (QCD). Since the exact form of confinement from QCD
is not known, one has to go for phenomenological models. The
phenomenological models are either non-relativistic quark models
(NRQM)~\cite{Gr77,Gr78,BCN81,GI85,SG85,SS92,BBH90} with a suitably
chosen potential, or relativistic models where the interaction is
treated perturbatively. NRQM have proved to be quite successful in
describing the hadronic properties. The Hamiltonian of these quark
models usually contains three main ingredients: the kinetic energy,
the confinement potential and an hyperfine interaction term, which
has often been taken as an effective one-gluon-exchange potential
(OGEP) \cite{RG75}. Other types of hyperfine interaction have been
introduced in the literature; from the non-relativistic reduction of
the t'Hooft interaction \cite{tH76,SVZ80}, termed as
Instanton-Induced Interaction (III), which has already been
successfully applied in several studies of the hadron spectra
~\cite{BBH90,Loe01,Me00,SS97,PH85,OT91,Kochelev:1985de}. The main
achievement of the III in hadron spectroscopy is the resolution of
the $U_A(1)$ problem, which leads to a good prediction of the masses
of $\eta$ and $\eta'$ mesons. The Goldstone-Boson-Exchange
interaction introduced by Glozman and Riska \cite{GR96} furnishes
another example of hyperfine interaction; it allows a good
description of the baryon spectrum and gives a correct ordering for
the positive and negative parity states. The model of Glozman and
Riska has however the major caveat to apply only to baryons and is
thus not
able to give an unified description of the spectrum of hadrons.\\

The successes of the NRQM in describing the spectrum of light
hadrons are somehow paradoxical, as light quarks should in principle
not obey a non-relativistic dynamics. This paradox has been avoided
in many works based on the constituent quark model by using for the
kinetic energy term of the hamiltonian a semi-relativistic or
relativistic expression (see, for example,\cite{SS97,OY99,BS00}).

In literature there are models which have tried to explain hadron
spectroscopy only with OGEP~\cite{Gr77,GI85} and some models with
only III~\cite{BBH90,SS97}, ignoring completely the OGEP. We feel
that it may be an exaggeration to eliminate OGEP completely for
light quarks. The OGEP still has to be present but with a smaller
strength consistent with the asymptotic freedom, since the III has
to vanish for heavy quarks.

In our present work, we have made use of the relativistic harmonic
model (RHM) \cite{KG83}. The RHM combined with OGEP has already been
used to calculate light hadrons masses, baryons magnetic moments,
leptonic decay widths and N-N scattering phase shifts
\cite{Vi93,Vi97,VP99}. Previously, by employing the RHM
~\cite{KG83,VP04} and the NRQM~\cite{BV05} along with III, the
ground state masses of pseudoscalar and vector mesons were
investigated. In both the cases the results showed that the
inclusion of III diminished the relative importance of OGEP for the
hyperfine splitting. One of the aims of the investigation was also
to test whether quark gluon coupling constant($\alpha_{s}$) can be
treated as a perturbative effect and to understand the role played
by the III in meson spectra~\cite{VP04}.

In view of the apparent success of RHM ~\cite{VP04} in the
description of $S$-wave spectra of mesons, we feel it is worthwhile
to apply it to the case of orbitally excited states. In our present
work  the full hamiltonian used has a Lorentz scalar plus vector
confinement potential, along with central and non-central
(spin-orbit and tensor) terms of OGEP and III. In addition, III also
has antisymmetric spin-orbit term proportional to
$\vec{L}\cdot\vec{\triangle}$ where $\vec{\triangle}$ is defined in
terms of the Pauli matrices as $\frac{1}{2}(\vec{\sigma}_{1}-
\vec{\sigma}_{2})$. The full discussion of the hamiltonian is given
in section II. We also discuss the parameters involved in our model
in section ~\ref{sec:parameters}. The results of the calculation are
presented in section ~\ref{sec:results} and the conclusions are
given in section ~\ref{sec:conclusions}.

In our present work, we have computed the masses of the singlet and
triplet light P wave mesons by including the III as a short-range
nonperturbative gluon effect in addition to the perturbative
conventional OGEP derived from QCD. This will allow much better
understanding of the $P$-wave meson spectroscopy, where some of the
$q\bar{q}$ quark model assignments of the known mesons are still
controversial. We hope it will also allow us a better understanding
of the production properties of the $P$-wave mesons. In literature
there are numerous attempts to understand the $P$-wave meson
spectroscopy. The reference can be found in the review ~\cite{WL91}.
One of the aims of this study is to determine explicitly the role
played by instantons in meson spectra, when used in the framework of
the RHM and to compare the effects of III to that of OGEP.

\section{The relativistic harmonic model}
\label{sec:relmodel}

In the RHM~\cite{KG83}, quarks in a hadron are confined through
the action of a Lorentz scalar plus a vector harmonic oscillator
potential
\begin{eqnarray}
\label{(1)} V_{\rm conf}(r) = \frac{1}{2}(1+\gamma_0)\ A^2 r^2 + M
\end{eqnarray}
where $\gamma_0$ is the Dirac matrix
\begin{eqnarray}
\label{(2)}\gamma _0=\left(
\begin{array}{cc}
1 & 0 \\
0 & -1
\end{array}
\right)
\end{eqnarray}
$M$ is the quark mass parameter and $A^2$ the confinement strength.
They have a different value for each quark flavour. In the RHM, the
confined single quark wave function $\Psi$ is given by
\begin{eqnarray}
\label{(3)} \Psi =N\ \left(
\begin{array}{c}
\Phi \\
\frac{\boldsymbol{\sigma}\cdot\boldsymbol{P}}{E+M} \ \Phi
\end{array}
\right)
\end{eqnarray}
with the normalization
\begin{eqnarray}
\label{(4)} N=\sqrt{\frac{2(E+M)}{3E+M}},
\end{eqnarray}
$E$ is an eigenvalue of the single particle Dirac equation with the
interaction potential given by (\ref{(1)}). We can perform a unitary
transformation to eliminate the lower component of $\Psi$ such that
\begin{eqnarray}
\label{(5)} U\ \Psi = \left(
                        \begin{array}{c}
                          \Phi \\
                          0 \\
                        \end{array}
                      \right)
\end{eqnarray}
where $U$ is given by
\begin{eqnarray}
\label{(6)} U = \frac {1}{N\ \left[ 1+\frac{P^2}{(E+M)^2}\right] }
\left(
\begin{array}{cc}
1 & \frac{{\bf \sigma .P}}{E+M} \\ -\frac{{\bf \sigma .P}}{E+M} &
1
\end{array}
\right)
\end{eqnarray}
Here $U$ is a momentum and eigenvalue $E$ dependent transformation
operator. With this transformation, the upper component $\Phi$
satisfies the equation
\begin{eqnarray}
\label{(7)} \left[ \frac{P^2}{E+M}+ A^2 r^2\right] \ \Phi =(E-M)\
\Phi
\end{eqnarray}
which is like the three dimensional harmonic oscillator equation
with an energy dependent parameter $\Omega_n^2$
\begin{eqnarray}
\label{(9)}\Omega _n^{2}=A \ (E_{n}+ M )^{\frac{1}{2}}
\end{eqnarray}
 The eigenvalue of (\ref{(7)}) is thus given by
\begin{eqnarray}
\label{(8)} E_n^2=M^2+(2n+1)\ \Omega_n^{2}
\end{eqnarray}

Note that Eq.(\ref{(7)}) can also be derived by eliminating the
lower component of the wave function, using a Foldy-Wouthuysen
transformation, as it has been done in \cite{KG83}.

The total energy of the hadron is obtained by adding the individual
contributions of the quarks. The spurious center of mass (CM) is
corrected \cite{GS57} by using intrinsic operators for the $\sum_i
r_i^2$ and $\sum_i \nabla_i^2$ terms appearing in the Hamiltonian.
This amounts to just subtracting the CM motion zero contribution
from the $E^2$ expression.

We come now to the description of the quark-antiquark potential; it
is given by the sum of  OGEP and III potential
\begin{eqnarray}
\label{(10)} V_q(r_{ij}\ )=V_{\rm OGEP}(r_{ij})+V_{\rm III}(r_{ij})
,
\end{eqnarray}
with $r_{ij}$ the inter-quark distance. Among the several versions
of the OGEP, we have used the following one, first derived in
\cite{RG75} from the QCD Lagrangian in the non-relativistic limit
\begin{eqnarray}
\label{(11)}V_{\rm OGEP}(r_{ij}) & = & \frac{\alpha_s}4
{\bm\lambda}_i\cdot{\bm\lambda}_j\\
& & \times \left[ \frac 1{r_{ij}}-\frac \pi {M_i M_j}\ (1+\frac 23\
{\bf \sigma }_i\cdot {\bf\sigma }_j)\ \delta
(r_{ij})\right]\nonumber
\end{eqnarray}
where the first term is the residual Coulomb energy and the
second-term the chromo-magnetic interaction leading to the hyperfine
splittings. Here, $\lambda_{i}$ is the generator of the color SU(3)
group for the $i^\mathrm{th}$ quark, $\sigma_{i}$ is the Pauli spin
operator.

To model the III, we have used the form given in
~\cite{BBH90,SS97}:
\begin{eqnarray}
 V_\mathrm{III}(r_{ij})
 = \left\{
          \begin{array}{ll}
           - 8 g  \delta(r_{ij}) \delta _{S,0} \delta _{L,0} , &
           I=1,  \\
           - 8 g' \delta(r_{ij}) \delta _{S,0} \delta_{L,0} , &
           I=1/2, \\
            8 \left(
                     \begin{array}{cc}
                      g & \sqrt{2}g' \\
                      \sqrt{2}g' & 0
                     \end{array}
               \right) \delta(r_{ij}) \delta _{S,0} \delta_{L,0} , &
           I=0.
          \end{array}
   \right.
 \label{eq:V-III}
 \end{eqnarray}

In the above expression, $S, L, I$ are respectively the spin, the
relative orbital angular momentum and the isospin of the system. The
$g$ and $g'$ are dimensioned coupling constants. The Dirac delta
function appearing in (~\ref{eq:V-III}) needs to be regularized for
practical calculations. As in Ref.\cite{BBH90,SS97}, we have chosen
a Gaussian-like function
\begin{eqnarray}
\label{(13)} \delta(r_{ij}) \rightarrow \frac 1{(\Lambda \sqrt{\pi
})^3}\exp [-\frac{r_{ij}^2}{\Lambda ^2}]\
\end{eqnarray}

The non-central part of the OGEP consists of the tensor term $
V_\mathrm{OGEP}^\mathrm{T}(\vec{r}_{ij})$ and the spin-orbit
interaction $V_\mathrm{OGEP}^\mathrm {SO} (\vec{r}_{ij})$. There are
several versions of tensor term in literature~\cite{WL91}. We use
the expression derived in~\cite{RG75} from the QCD lagrangian in the
non-relativistic limit and used subsequently by many authors (for
example  ~\cite{AB89,AV95})
\begin{eqnarray}
V_\mathrm{OGEP}^\mathrm{T}(\vec r_{ij}) =
-\frac{\alpha_s}{4}{\bm\lambda}_i \cdot {\bm\lambda}_j
\left[\frac{1}{4  M_{i}M_{j}} \frac{1}{r_{ij}^{3}}\right]
\hat{S}_{ij} \label{eq:V-tOGEP}
\end{eqnarray}
where $\hat{S}_{ij} =
3(\vec{\sigma}_{i}\cdot\hat{r})(\vec{\sigma}_{j}\cdot\hat{r})-\vec{\sigma}_{i}\cdot\vec{\sigma}_{j}$.
The tensor potential is a scalar which is obtained by contracting
two second rank tensors. Here, $\hat{r}=\hat{r}_{i}-\hat{r}_{j}$ is
the unit vector in the direction of $\hat{r}$. Note that in the
presence of the tensor interaction, $ \vec{L}$ is no longer a good
quantum number.

The spin-orbit (SO) interaction of the OGEP is given by
\begin{eqnarray}
V_\mathrm{OGEP}^\mathrm {SO} (\vec r_{ij}) & = &
-\frac{\alpha_s}{4} {\bm\lambda}_i \cdot {\bm\lambda}_j \\
& \times & \left[\frac{3}{8 M_{i}M_{j}} \frac{1}{r_{ij}^{3}}
\left({\vec r_{ij}}\times {\vec P_{ij}} \right)
\cdot\left({\vec\sigma}_i\ + {\vec\sigma}_j\right)\right]\nonumber
\label{eq:V-lsOGEP}
\end{eqnarray}
where the angular momentum is defined as usual in terms of
relative position $\vec r_{ij}$ and the relative momentum $\vec
P_{ij}$. Unlike the tensor force, the spin-orbit force does not
mix states of different $\vec{L}$, since $ L^{2}$ commutes with
$\vec{L}\cdot\vec{S}$, $\vec{L}$ is still a constant of motion,
but $L_{z}$ is not.

The spin-orbit term of III is (see Refs.~\cite{BBH90,SS97}) given by
\begin{eqnarray}
V^\mathrm{SO}_\mathrm{III}(\vec r_{ij}) =
V_\mathrm{LS}(r_{ij})\vec{L}\cdot\vec{S}+ V_{\mathrm
L\bigtriangleup}(r_{ij})\vec{L}\cdot\vec{\triangle}
\label{eq:V-lsIII}
\end{eqnarray}
The first term in Eq.~(\ref{eq:V-lsIII}) is the traditional
symmetric spin-orbit term proportional to the operator
$\vec{L}\cdot\vec{S}$. The other term is the anti-symmetric
spin-orbit term proportional to $\vec{L}\cdot\vec{\triangle}$, where
$\vec{\triangle}=\frac{1}{2}(\vec{\sigma}_{1}-\vec{\sigma}_{2})$.
The radial functions of Eq.~(\ref{eq:V-lsIII}) are expressed as
\begin{eqnarray}
 V^\mathrm{LS}_\mathrm{III}(r_{ij})=  \left(
\frac{1}{M_{i}^{2}}+\frac{1}{M_{j}^{2}}\right)\sum_{k=1}^{2}\kappa_{k}\frac{\exp(-
r^{2}_{ij}/\eta_{k}^{2})} {(\eta_{k}\sqrt{\pi})^{3}}\nonumber\\+
\frac{1}{M_{i}M_{j}}\sum_{k=3}^{4}\kappa_{k}\frac{\exp(-r^{2}_{ij}/\eta_{k-2}^{2})}{(\eta_{k-2}\sqrt{\pi})^{3}}
\label{eq:V-iiiLS}
\end{eqnarray}
and
\begin{eqnarray}
V^\mathrm{L\Delta}_\mathrm{III}(r_{ij})=  \left(
\frac{1}{M_{i}^{2}}-\frac{1}{M_{j}^{2}}\right)\sum_{k=5}^{6}\kappa_{k}\frac{\exp(-
r^{2}_{ij}/\eta_{k-4}^{2})} {(\eta_{k-4}\sqrt{\pi})^{3}}
\label{eq:V-iiiDEL}
\end{eqnarray}

The tensor term of III is
\begin{eqnarray}
V^\mathrm{T}_\mathrm{III}(\vec r_{ij})=\frac{\hat
S_{ij}}{M_{i}M_{j}}\sum_{k=7}^{8}\kappa_{k}\frac{\exp(-r^{2}_{ij}/\eta_{k-4}^{2})}
{(\eta_{k-4}\sqrt{\pi})^{3}} \label{eq:V-iiiTEN}
\end{eqnarray}
The term $V_\mathrm{LS}(r_{ij})$ is responsible for the splitting of
the $^{3}L_{J}$ states with $J = L-1, L, L+1$. The term
$V_{L\bigtriangleup}(r_{ij})$ couples states $^{1}L_{J=L}$ and
$^{3}L_{J=L}$ and due to mass dependence this term is inoperative
when the quarks are identical. It is to be noted that the III and
the OGEP have the same spin dependence except for the
$V_{L\bigtriangleup}$ term.

\section{Fitting Procedure}
\label{sec:parameters}

The parameters of the RHM are the masses of the quarks, $M_u = M_d$
and $M_s$, the respective confinement strengths, $A_u^2 = A_d^2$,
$A_{s}^2$, and the oscillator size parameter $b_{n}\ (=
1/{\Omega_{n} }$). They have been chosen to reproduce various
nucleon's properties: the root mean square charge radius, the
magnetic moment and the ratio of the axial coupling to the vector
coupling~\cite{KG83}. The confinement strength $A_{u,d}$ is fixed by
the stability condition for the nucleon mass against the variation
of the size parameter $b_{n}$
\begin{eqnarray}
\label{(14)}\frac \partial {\partial b_{n}}\langle N\left| H\right|
N\rangle = 0.
\end{eqnarray}
The value of $b_{n}$ for S wave was 0.77 fm ~\cite{VP04} and for P
wave $b_{n}$ is 0.7 fm. It is to be noted that $b_{n}$ is state
dependent. The parameters associated with the strange quark $M_s$
and $A_s^2$ have been fitted in order to reproduce the magnetic
moments of the strange baryons, according to the procedure described
in~\cite{GK87}. The coupling constant $\alpha_s$ of OGEP is fixed
from S wave meson spectroscopy~\cite{VP04}. The value of $\alpha_s$
turns out to be 0.2 for P wave mesons, which is compatible with the
perturbative treatment. The parameters of central part of III are
$g$, $g^\prime$ and $\Lambda$, the strength and the range of the
interaction of III, which were fitted to the experimental masses of
$\pi$ and $K$ mesons~ \cite{BV05}.

Among the non-central parts of the potential, the hyperfine terms of
III has 12 additional strength and size parameters $\kappa$ and
$\eta$ (in Eq.~\ref{eq:V-iiiLS}-\ref{eq:V-iiiTEN}) respectively. We
are able to reproduce the light $P$ wave meson masses with all
$\eta$ and $\kappa_1$ to $\kappa_6$ parameters held fixed and by
varying only the $\kappa_7$ and $\kappa_8$ parameters. The values of
$\kappa_7$ and $\kappa_{8}$ are listed in
Table~\ref{Tab:parameters-2}. It is to be noted that for each
category of meson nonet, $\kappa_7$ is held fixed and only the
$\kappa_{8}$ is varied. The non-central terms of OGEP are
attractive, whereas the strengths of the interaction of III (i.e.,
$\kappa$) can have both positive and negative values~\cite{SS97}. We
are led to the conclusion that inclusion of III in the formalism is
essential. This also enables us to bring down the value of
$\alpha_s$ to 0.2.
\begin{table}
\caption{\label{Tab:parameters-1} Values of parameters used in our
model.}
\begin{ruledtabular}
\begin{tabular}{r|l}

 $b_{n}$              & 0.7 fm                           \\
 $M_\mathrm{u,d}$ & 315 MeV                          \\
 $M_\mathrm{s}$   & 450 MeV                          \\
 $\alpha_s$       & 0.2                               \\

 $A_\mathrm{u}^{2}=A_\mathrm{d}^{2}$  & 3754.7 MeV fm$^{-2}$ \\
 $A_\mathrm{s}^{2}$   & 3367.45 MeV fm$^{-2}$     \\
 $g$              & 0.0847 $\times$ 10$^{-4}$ MeV$^{-2}$    \\
 $g^\prime$             & 0.0535 $\times$ 10$^{-4}$ MeV$^{-2}$    \\
 $\Lambda$        & 0.35 fm                           \\
 $\eta_{1}$     & 0.194 fm \\
 $\eta_{2}$      & 0.294 fm \\
 $\eta_{3}$     & 0.112 fm \\
 $\eta_{4}$     & 0.501 fm \\

 $\kappa_{1}$     & -2.213 \\
 $\kappa_{2}$     & 0.191 \\
 $\kappa_{3}$     & -2.13 \\
 $\kappa_{4}$     & 2.651 \\
 $\kappa_{5}$     & 20.84 \\
 $\kappa_{6}$    & 26.43 \\
\end{tabular}
\end{ruledtabular}
\end{table}

\begin{table}
\caption{\label{Tab:parameters-2} Values of $\kappa_7$ and
$\kappa_{8}$ parameters used in our model.}
\begin{ruledtabular}
\begin{tabular}{c|ccc}
$N^{2S+1}L_{J}$& Meson &  $\kappa_{7}$& $\kappa_{8}$\\
\hline\\
      $1^{3}P_{0}$& $f_{0}\mathrm{(600)} ({\rm or~}\sigma)$      & -31.63 & 23.92   \\
      $1^{3}P_{0}$& $f_{0}\mathrm{(980)}$      & -31.63 & 17.35    \\
      $1^{3}P_{0}$& $a_{0}\mathrm{(980)}$      & -31.63 & 6.81    \\
      $1^{3}P_{0}$& $K_{0}^{\ast}\mathrm{(1430)}$  & -31.63 & -21.19   \\
      $1^{3}P_{0}$& $a_{0}\mathrm{(1450)}$      & -31.63& -87.45   \\
      $1^{3}P_{0}$& $f_{0}\mathrm{(1500)}$  & -31.63& -47.26   \\
      \hline\\
      $1^{3}P_{1}$& $a_{1}\mathrm{(1260)}$   & 28.8 & 1.65       \\
      $1^{3}P_{1}$& $K_{1}\mathrm{(1270)}$   & 28.8 & -1.67        \\
      $1^{3}P_{1}$& $f_{1}\mathrm{(1285)}$   & 28.8 & -10.32       \\
      $1^{3}P_{1}$& $f_{1}\mathrm{(1420)}$   & 28.8 & 18.59 \\
      \hline\\
      $1^{3}P_{2}$& $f_{2}\mathrm{(1270)}$      & 22.84 & -37.95  \\
      $1^{3}P_{2}$& $a_{2}\mathrm{(1320)}$      & 22.84 & 25.52  \\
      $1^{3}P_{2}$& $K_{2}^{\ast}\mathrm{(1430)}$  & 22.84 & 35.19 \\
      $1^{3}P_{2}$& $f_{2}^\prime\mathrm{(1525)}$     & 22.84 & 47.23\\

\end{tabular}
\end{ruledtabular}
\end{table}

The $q\bar{q}$ wave function for each meson is expressed in terms of
oscillator wave functions corresponding to the CM and relative
co-ordinates. The oscillator quantum number for the CM wave
functions are restricted to $N_{cm} = 0$. The Hilbert space of
relative wave functions is truncated at radial quantum number $n =
2$. The Hamiltonian matrix is constructed for each meson separately
in the basis states of $\vert N_{cm}=0, L_{cm}=0;
N^{2S+1}L_{J}\rangle$ and diagonalised.

\section{Results and discussion}
\label{sec:results}  The masses of the singlet and triplet P-wave
mesons after diagonalisation in harmonic oscillator basis with
$n_{max}$=2 are listed in Table~\ref{Tab:sing-P} and
\ref{Tab:trip-P} respectively.

\begin{table}
\caption{\label{Tab:sing-P} The pseudo-vector meson masses (in
MeV).}
\begin{ruledtabular}
\begin{tabular}{c|cccc}
   Meson     & $b_{1}\mathrm{(1235)}$   & $h^\prime_1\mathrm{(1380)}$ & $K_1\mathrm{(1400)}$ \\

\hline\\
Experimental Mass & 1229.5$\pm$3.2 & 1386$\pm$19  & 1402$\pm$7  \\
Calculated Mass   &1229.22         &  1386.42    & 1408.34  \\

\end{tabular}
\end{ruledtabular}
\end{table}

\begin{table}
\caption{\label{Tab:trip-P} The triplet meson masses (in MeV).}
\begin{ruledtabular}
\begin{tabular}{c|ccc}
$N^{2S+1}L_{J}$& Meson  & Experimental Mass & Calculated Mass    \\
\hline\\
      $1^{3}P_{0}$& $f_{0}\mathrm{(600)} ({\rm or} \sigma)$ & 400-1200 & 703.54   \\
      $1^{3}P_{0}$& $f_{0}\mathrm{(980)}$ & 980 $\pm$ 10 & 989.72      \\
      $1^{3}P_{0}$& $a_{0}\mathrm{(980)}$ & 984.7$\pm$1.2 & 986.84     \\
      $1^{3}P_{0}$& $K_{0}^{\ast}\mathrm{(1430)}$ & 1412$\pm$6 &1412.07   \\
      $1^{3}P_{0}$& $a_{0}\mathrm{(1450)}$ & 1474$\pm$19 & 1471.77     \\
      $1^{3}P_{0}$& $f_{0}\mathrm{(1500)}$ & 1507$\pm$5 & 1507.38      \\
      \hline\\
      $1^{3}P_{1}$& $a_{1}\mathrm{(1260)}$   & 1230$\pm$40         & 1229.54    \\
      $1^{3}P_{1}$& $K_{1}\mathrm{(1270)}$   & 1273$\pm$7          & 1272.14    \\
      $1^{3}P_{1}$& $f_{1}\mathrm{(1285)}$   & 1281.8$\pm$0.6      & 1281.15    \\
      $1^{3}P_{1}$& $f_{1}\mathrm{(1420)}$   & 1426.3$\pm$0.9      &1426.74     \\
      \hline\\
      $1^{3}P_{2}$& $f_{2}\mathrm{(1270)}$      & 1275.4$\pm1.2$      & 1276.83      \\
      $1^{3}P_{2}$& $a_{2}\mathrm{(1320)}$      & 1318$\pm$0.6        & 1317.80      \\
      $1^{3}P_{2}$& $K_{2}^{\ast}\mathrm{(1430)}$  & 1425.6$\pm$1.5      &1423.61       \\
      $1^{3}P_{2}$& $f_{2}^\prime\mathrm{(1525)}$     & 1525$\pm$5          & 1525.9      \\
\end{tabular}
\end{ruledtabular}
\end{table}

Our results indicates that only the OGEP hyperfine interaction is
not sufficient to reproduce the masses of the mesons. If OGEP is
taken as the only source of hyperfine interaction, the value of
$\alpha_s$ necessary to reproduce the hadrons spectrum is
generally much larger than one; this leads to a large spin-orbit
interaction, which destroys the overall fit to the spectrum. The
inclusion of III will diminish the relative importance of OGEP for
the hyperfine splittings.  The important role played by the III in
reproducing the masses of these mesons (as shown in
Table~\ref{Tab:trip-P}) can be gauged by examining the Table
~\ref{Tab:mass2} where the masses of the scalar mesons calculated
after switching off the III in the full Hamiltonian are tabulated.
\begin{table}
\caption{\label{Tab:mass2} The masses of scalar mesons (in MeV)
after diagonalisation without III}
\begin{ruledtabular}
\begin{tabular}{c|ccccc}
 Meson   & Experimental mass & Calculated  Mass\\
\hline\\
 $a_{0}$(1450) &1474$\pm$19  &1002.65\\
 $ K_{0}^{\ast}$(1430)& 1412  $\pm$6 & 1157.45\\
 $ f_{0}$(1500)& 1507$\pm$5 & 1259.84\\
\end{tabular}
\end{ruledtabular}
\end{table}
This is because the tensor and spin-orbit terms of OGEP are
attractive and hence bring down the masses of the triplet state.
Also, it is important to note that, the spin-orbit terms of III
are very weak ~\cite{SS97}. The dominant contribution to the
splitting of masses of axial vector mesons comes from the tensor
term of III, which involves the parameters $\kappa_{7}$ and
$\kappa_{8}$. It was necessary to tune $\kappa_{7}$ and
$\kappa_{8}$ parameters so as to get a reasonably good agreement
with the experimental masses. Hence, in our model we have only two
free parameters $\kappa_{7}$ and $\kappa_{8}$. Also, the results
indicate that tensor part of III is crucial and plays a dominant
role in explaining the masses of P wave mesons which is an
important result of our investigation. The attractive or repulsive
nature of III being governed by the sign of the $\kappa$. Thus by
tuning the $\kappa$ parameters appropriately, we are able to
reproduce the meson masses in our model.

\subsection{Pseudo-vector meson nonet ($1^{1}P_{1}$)}

We have investigated three mesons of the $1^{1}P_{1}$ pseudovector
meson nonet with $J ^{PC}=1^{+-}$, namely, $b_{1}\mathrm{(1235)}$,
$h_{1}^\prime\mathrm{(1380)}$, $K_{1}\mathrm{(1400)}$~\cite{PDG}. It
may be pointed out here that there is no contribution from the III
for the singlet states except for the $^{1}P_{1}$ state in the
K-sector. In the K-sector, the singlet P state receives a
significant repulsive contribution of 129.75 MeV from the off
diagonal matrix element (ME) of
$\langle^{3}P_{1}|V_{L\Delta}|^{1}P_{1} \rangle$. The masses
calculated by our model are tabulated against the corresponding
masses~\cite{PDG} in Table~\ref{Tab:sing-P}

\subsection{Scalar meson nonet($1^{3}P_{0}$)}

The spectrum of the scalar meson nonet is very large and the actual
number of resonances in the region of 1-2 GeV far exceeds the number
of states which the conventional quark models can accommodate.
Several of these states, however, have been interpreted as exotic
mesons. It is well known that a $q\bar{q}$ meson with orbital
angular momentum $l$ and total spin $s$ must have parity
$P=(-1)^{l+1}$ and charge conjugation quantum number $C=(-1)^{l+s}$.
On this basis, we define an exotic meson to be one which does not
have the above spectroscopic configurations. Thus a resonance with
$J^{PC}=0^{--}, 0^{+-}, 1^{-+}, 2^{+-} \cdots$ are exotic. Such
states could be a gluonic excitation such as a hybrid ($q\bar{q}g$)
or glue ball ($ 2g,3g\cdots $) or a multi quark
state($q\bar{q}q\bar{q}$).

The particle data group(PDG) lists isoscalar states, the
$a_{0}$(980) and $a_{0}$(1450)~\cite{PDG} having masses of $984.7
\pm 1.2$ MeV and $1450 \pm 40$ MeV respectively. Theories based on
chiral sigma models with three flavors \cite{WU04} suggest that
$a_{0}$(980) would form a scalar nonet. The scalar
$K^{\ast}_{0}(1430)$ is well established. Several groups have
claimed different isoscalar structures close to 1500 MeV
\cite{SA94,DVB95}. In this work, we focus our attention only on the
non-exotic scalar mesons with $J^{PC}=0^{++}$ and assigned as
$f_{0}\mathrm{(600)}$, $f_{0}\mathrm{(980)}$, $a_{0}\mathrm{(980)}$,
$K_{0}^{\ast}\mathrm{(1430)}$, $a_{0}\mathrm{(1450)}$ and
$f_{0}\mathrm{(1500)}$~\cite{PDG}.The isosinglet
$f_{0}\mathrm{(980)}$and and isotriplet $a_{0}$(980) lower than 1
GeV are considered as a chiral partner $(\sigma$ nonet) of the
pseudo scalar ($(\pi$ nonet) in SU(3) chiral symmetry.The light
scalar mesons with masses less than 1 GeV are not considered as a
conventional L=1 scalar nonet by many authors.

\subsection{Axial vector meson nonet ($1^{3}P_{1}$)}

In our model, for axial vector mesons, the tensor and
$\vec{L}\cdot\vec{S}$ parts of OGEP and III have opposite signs. The
contributions due to tensor terms are repulsive, whereas those due
to  $\vec{L}\cdot\vec{S}$ terms are attractive. As the OGEP has the
same strength parameter for these terms, the contribution of the
hyperfine interaction terms of OGEP is negligible whereas, due to
the different strength parameters $\kappa_i$, the corresponding
terms of III contribute differently. Besides, the contribution of
III to the masses is also significant because of the different
radial form of tensor and spin-orbit terms. We have treated
$\kappa_{i}$ as free parameters so as to reproduce the masses of
$a_{1}\mathrm{(1260)}$, $K_{1}\mathrm{(1270)}$,
$f_{1}\mathrm{(1285)}$ and $f_{1}\mathrm{(1420)}$. However, it
should be noted that the $a_{1}$(1260), with $I = 1$ has a
significant width of ~ 400 MeV and has a dominant decay channel
$a_{1}\rightarrow\rho\pi$. This property makes the determination of
its mass difficult. The QCD sum rules ~\cite{LJ85} produce a mass of
$1150 \pm 40$ MeV. According to Bowler~\cite{BO86}, the $a_{1}$ mass
and width are safely within the ranges $\simeq 1235 \pm 40$ MeV and
$400 \pm 100$ MeV respectively. These values are in agreement with
those currently adopted by PDG~\cite{PDG}, i.e., mass of $1230 \pm
40$MeV and width 250 MeV to 600 MeV. In the K-sector, we have fitted
to $K_{1}$(1270). The contribution from the matrix element (ME)
$\langle1^{1}P_{1}|V_{L\Delta}|1^{3}P_{1}\rangle$ has been found to
be significant. PDG cite two $f_1$ meson states \cite{PDG} with
$J^{PC}=1^{++}$, namely, $f_{1}$(1285) and $f_{1}$(1420). There has
been considerable discussion on the quark structure of these mesons
\cite{BUR97}. We have been able to fit the masses reasonably well as
shown in Table~ \ref{Tab:trip-P}.

\subsection{Tensor meson nonet ($1^{3}P_{2}$)}

We consider some of the well established members of the tensor meson
nonets, with $J^{PC}=2^{++}$, i.e., $f_{2}\mathrm{(1270)}$,
$a_{2}\mathrm{(1320)}$, $K_{2}^{\ast}\mathrm{(1430)}$ and
$f_{2}^\prime\mathrm{(1525)}$. The contributions due to tensor and
$\vec{L}\cdot\vec{S}$ terms of OGEP and III bear opposite signs. The
tensor potential is attractive whereas $\vec{L}\cdot\vec{S}$ part is
repulsive. However, the off diagonal tensor ME
$\langle{3}P_{2}|V_{OGEP}^{\mathrm T}|^{3}F_{2}\rangle$ is strongly
repulsive. In our model the mass difference between $f_{1}$ and
$f_{2}^\prime$ essentially comes from the off diagonal ME of tensor
potential of OGEP and III.

In literature some more $J^{PC}= 2^{++}$ states like
$f_{2}(1520)$, $f_{2}(1810)$, $f_{2}(2010)$, $f_{2}(2340)$ have
been considered. Of these, $f_{2}(1810)$ is likely to be the
$2^{3}P_{2}$ state \cite{BUR97}. Our model prediction for
$f_{2}(1810)$ is 1724.52 MeV.

\section{Conclusions}
\label{sec:conclusions}

The mass spectrum of P wave mesons is considered in the frame work
of  RHM with the conventional OGEP and by including III.  The
inclusion of III consequently diminish the relative importance of
OGEP. The III also restricts $\alpha_{s}$ to be 0.2 and thus
justifying the perturbative truncation of multi gluon exchanges. The
near mass degeneracy of the experimentally established iso-doublet
states of the scalar and tensor meson nonets $K_{0}^{\ast}$ and
$K_{2}^{\ast}$ could be accounted by the off diagonal tensor ME of
OGEP and III. The simultaneous mass degeneracy of the pseudo-vector
$K_{1B}$ and axial vector $K_{1A}$ which mix to give physical
$K_{1}(1270)$ and $K_{1}(1400)$ states observed experimentally could
be accounted for by the anti-symmetric spin-orbit term $V_{L\Delta}$
of III. As we have shown, RHM  with OGEP and III provides a quite
good description of the pseudo-vector, scalar, axial vector and
tensor P-wave mesons with the same constituent quark masses,
oscillator size and OGEP strength $\alpha_{s}$. The calculated
masses of the P wave mesons are in good agreement with the
experimental masses.

\begin{acknowledgements}

One of the authors ($APM$) is grateful to, DST, India, for granting
the JRF. The other author (KBV) is thankful to DST (Sanction no.
SR/S2/HEP-14/2006). This work is supported in part by TWAS-UNESCO
program of ICTP, Italy and the Knowledge Innovation Project of
Chinese Academy of Sciences under contract No. KJCX2-YW-W10.
\end{acknowledgements}

\end{document}